# A One-Dimensional Model for Investigating Scale-separated Approaches to the Interaction of Oceanic Internal Waves

Kurt L. Polzin,[a] Yuri V. Lvov[b]

[a] *Woods Hole Oceanographic Institution, Woods Hole, Massachusetts*
[b] *Department of Mathematical Sciences, Rensselaer Polytechnic Institute, Troy, NY*

*Corresponding author*: Kurt L. Polzin, kpolzin@whoi.edu

ABSTRACT: High-frequency wave propagation in near-inertial wave shear has, for four decades, been considered fundamental in setting the spectral character of the oceanic internal wave continuum and for transporting energy to wave-breaking. We compare idealized ray tracing numerical results with metrics derived using a wave turbulence derivation for the kinetic equation and a path integral to study such scale-separated interactions. These diagnostics include the mean drift in vertical wavenumber, the dispersion about that mean drift, time lagged correlation estimates of wavenumber and phase locking of the wave packets with the background. At small inertial wave amplitudes, all three provide consistent descriptions for the mean drift of wavepackets in the spectral domain and dispersion about that mean drift. Extrapolating our results to the background internal wavefield over-predicts downscale energy transports by an order of magnitude. At oceanic amplitudes, however, the numerics support diminished transport and dispersion that coincide with the mean drift time scale becoming similar to the lagged correlation time scale. We parse this as the transition to a non-Markovian process. Despite this decrease, numerical estimates of downscale energy transfer are still significantly larger than oceanic derived metrics. We argue that residual differences between the ray-path formulation and observations result from an unwarranted discard of Bragg scattering resonances, similar in manner to fourth-order cumulants providing an eddy damping term in 3-D turbulence. Our results support replacing the long standing interpretive paradigm of extreme scale-separated interactions with a more nuanced slate of 'local' interactions in the kinetic equation.

# 1. Introduction

In a companion paper concerning oceanic internal wave interactions (Lvov and Polzin, hereafter LP) we present two parallel derivations of spectral transport equations in an extreme scale-separated limit; one based upon traditional wave turbulence theory techniques and one based upon ray tracing. Both are generalized diffusion (Fokker-Planck) equations. However, they differ in that the ray-tracing version contains an explicit mean drift term. Evaluation of the extreme scale-separated transports for the vertical wavenumber - vertical wavenumber component of the diffusivity tensor leads to an immediate contradiction. The kinetic equation predicts no transport. In contrast, the related vertical wavenumber drift term of the ray tracing transport equation returns a downscale transport that is an order of magnitude larger than metrics of ocean mixing known as 'Finescale Parameterizations' (Polzin et al. 2014)! The difference between these theoretical predictions and what is required by our observational knowledge of sources and sinks is such that we (Polzin and Lvov 2017; Dematteis et al. 2022) refer to this growing set of contradictions as the 'Oceanic Ultraviolet Catastrophe': despite a tendency of oceanic spectral slopes in vertical wavenumber and frequency coinciding with stationary states of the Fokker-Planck equation (Polzin and Lvov 2011), there is extreme dissonance when one tries to rationalize the distribution of sources and sinks using the dynamical processes encapsulated in that Fokker-Planck equation.

This work follows analysis of the internal wave kinetic equation presented in Dematteis and Lvov (2021) and Dematteis et al. (2022). That work plays off a scale invariant stationary state of $n(\mathbf{p}) \propto k^{-3.69} m^0$ in the *non-rotating* limit, in which $n(\mathbf{p})$ is waveaction spectral density and $\mathbf{p}$ is a three dimensional wavenumber with vertical component $m$ and horizontal wavenumber magnitude $k$. That work formally introduces a 'cut' in the spectral domain to separate 'local' from 'extreme scale-separated' dynamics and finds that 'local' dynamics dominate and support downscale transports in both horizontal and vertical wavenumber. Local transports are sensibly consistent in magnitude and direction with the Finescale Parameterization. Extreme scale-separated transports are an order of magnitude smaller.

The importance of local interactions presents itself as a fundamental departure from the wisdom originating four decades ago that relates to the central role of 'Induced Diffusion' in constructing dynamical balances for the oceanic internal wavefield (McComas and Bretherton 1977; McComas and Müller 1981b; Müller et al. 1986). Induced Diffusion is the vertical-vertical component of the Fokker-Planck equation. The dominance of the vertical-vertical component is supported by a basic scale analysis presented in McComas and Bretherton (1977); Sun and Kunze (1999a). That wisdom is, however, grounded in significant mathematical tension: There is sentiment that the singularities of the integrand in the kinetic equation are most problematic in the vertical wavenumber domain (Lvov et al. 2010), but yet the background internal wavefield action density is independent of vertical wavenumber, so theoretical estimates of extreme scale-separated dynamics emanating from the kinetic equation assign small values to downscale transports in vertical wavenumber.

The wisdom of four decades ago comes with caveats (Holloway 1980, 1982; Müller et al. 1986) about the validity of the kinetic equation in this limit. Specifically, the concern is that the collisional cross sections represent a decidedly non-resonant quasi-coherent translation (i.e. advection or sweeping) of waves rather than the interactions that lead to energy transfer. The tools at that time were limited to resonant formulations and could not directly deal with the issue. In Polzin and Lvov (2017) we investigate resonance broadening using a canonical formulation in isopycnal coordinates to demonstrate that, indeed, at oceanic amplitudes, the bandwidth of the resonant manifold for extreme scale-separated interactions is proportional to the rms Doppler shift at oceanic amplitudes. This is aphysical, but a frequency renormalization of the broadened kinetic equation does not alter the transport prediction.

The wisdom of four decades ago also contains an effort to execute a similar investigation of kinetic theory and ray tracing (Henyey and Pomphrey 1983; Henyey et al. 1984). That work introduces intuitive notions of an interaction time scale and a correlation time scale, with kinetic theory being valid when interaction time scales are much larger than the correlation time scales (see also section 5 of Müller et al. 1986). There is a consensus that this time scale separation is problematic for the oceanic internal wavefield, with an underlying assumption that both are relatable to the resonant process. In this work, and in LP, we provide a mathematically grounded basis for 'interaction' and 'correlation' time scales by decomposing a wave-packet's excursions in phase space into an ensemble mean and perturbation. In doing so the ensemble mean drift in phase space relates to the interaction time scale and is a resonant process. The correlation time scale stems from a non-resonant process and relates to dispersion about the mean drift. The tension between interaction and correlation time scales appears concretely as a Markov process. But it comes at the cost of an explicit drift term in the ensemble mean transport equation which, in turn, over predicts oceanic dissipation rates by an order of magnitude and requires further elucidation. Our results for spectral drift and dispersion differ from those presented in McComas and Bretherton (1977); Nazarenko et al. (2001) as those derivations of transport consider the trajectory of a single ray rather than an ensemble of packets.

Ray methods have also been used in a 'kitchen sink' manner in which test waves are traced in a background consistent with a spectrally filtered version of the Garrett and Munk frequency - vertical wavenumber spectrum (Henyey et al. (1986); Sun and Kunze (1999b); Ijichi and Hibiya (2017)). These kitchen sink applications are numerical evaluations without theoretical support. Thus, for example, downscale transport estimates are arrived at via an intuitive advective assessment of counting the rate at which wavepackets cross a spectral gate. The theoretical justification for that methodology only appears in LP. In these 'kitchen sink' studies, the scale separations between test wave and background are viewed as tunable parameters to arrive at downscale transport estimates that align with observations. The alignment requires that the background have similar scales as the wave-packet and creates a thematic issue for an asymptotic theory such as ray tracing. In LP we provide a derivation of ray tracing and the action spectral balance to elucidate that the internal wave problem has an explicit scale separation in *horizontal* wavenumber in addition to an implicit averaging over the packet extent. Those kitchen sink simulations are inconsistent with our idealized model. We are able, in the context of our idealized model, to articulate why this could be so and thereby provide a theoretical understanding of the 'kitchen sink' mechanics.

In this paper we investigate closures for the ray tracing model in an effort to understand such discrepancies in the context of a simple one-dimensional model. The background for this numerical model is presented in Section 2. The metrics for mean drift and dispersion in vertical wavenumber are presented in Section 3. Results focused upon closures for the ray-tracing transport equation are presented in Section 4. Using these numerics we are able to identify issues relating to a Markov approximation at oceanic amplitudes that addresses differences between the idealized model and kitchen sink numerics, but this does not reconcile the order of magnitude discrepancy between our idealized prediction and ocean observations. This leads us into an inquiry about the physics eliminated from the ray-tracing model that might serve to decrease the coupling between high-frequency waves and inertial shear in section 5. Energy exchanges in the model result from the accumulation of wave packets under a group velocity equals phase velocity resonance condition. Implied is the creation of spatially local anisotropic wavefields over a time scale that is long in comparison to the time scale associated with a Bragg scattering process. Such a scattering mechanism is initiated by inertial waves having twice the wavenumber of the high-frequency packet and the scattered waves have oppositely signed vertical wavenumber with a reverse polarization signature that directly translates into the opposite sign of energy transfer relative to the original wave packet's phase velocity - group velocity coupling. Interaction with these half wavelength inertial waves is not accounted for in the standard ray-tracing paradigm. We argue that it is possible to parse such a four wave interaction as an eddy damped quasi-normal closure. We summarize in section 6.

## 2. A Scale Invariant Model of Wave Refraction in Inertial Shear

*a. Ray Equations*

In this section we describe a one-dimensional numerical ray tracing model similar to that presented in Polzin and Lvov (2017). The model uses (1) to represent the evolution of high-frequency test waves having wavenumber $\mathbf{p} = (k, 0, m)$ along trajectories in vertical wavenumber - intrinsic frequency space as in figure 1 and in space-time as in figure 2. We refer the reader to LP for a derivation of the eikonal relations $\dot{\mathbf{p}} = -\nabla_{\mathbf{r}} \sigma(\mathbf{p}, \mathbf{r}); \dot{r} = \nabla_{\mathbf{p}} \sigma(\mathbf{p}, \mathbf{r})$ that define ray trajectories in the wavenumber $\mathbf{p} = (\mathbf{k}, m)$ and spatial $\mathbf{r} = (x, y, z)$ domains from an Eulerian frequency $\sigma(\mathbf{p}, \mathbf{r})$ rendered as eq. 3.17 in LP. The one-dimensional model equations are:

$$\begin{aligned} \dot{m} &= -k \partial_z \mathcal{U}(z,t) \,, \\ \dot{z} &= \partial_m \omega \,, \\ \omega^2 - f^2 &= k^2 N^2 / m^2 \,, \end{aligned} \quad (1)$$

where $k$ is horizontal wavenumber, aligned with a purely horizontal background inertial flow $\mathcal{U}(z,t)$, $m$ is vertical wavenumber, $\omega$ is intrinsic frequency, $f$ is the Coriolis frequency and $\dot{}$ indicates a time derivative.

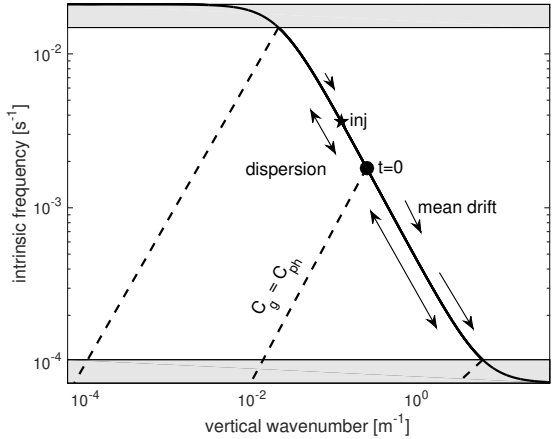

Figure 1. Schematic of trajectories in a vertical wavenumber - intrinsic frequency space for the one-dimensional model. high-frequency test waves are displaced along the solid black line with a mean drift to higher vertical wavenumber. The high-frequency test waves connect to the inertial field along the lower axis through a phase velocity equals group velocity resonance condition. The grey shading denotes the boundary of the model domain where the dispersion relation departs from the hydrostatic, nonrotating limit.

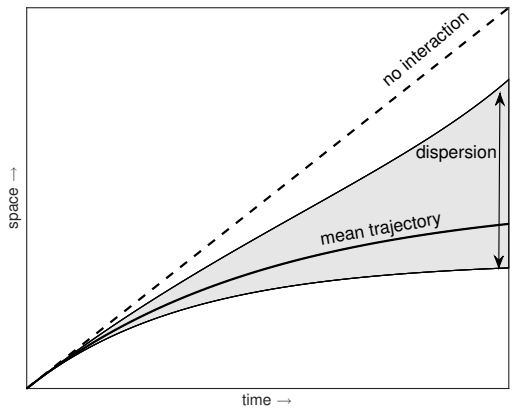

Figure 2. Schematic of the spatial - temporal evolution of the one-dimensional model. high-frequency test waves are displaced from the dashed line by interaction with the inertial field with dispersion (grey shading) about a mean drift towards smaller group velocity.

*b. Model Formulation*

The background (2) and (3) is comprised as a field of randomly phased inertial oscillations having no horizontal structure and a white vertical shear spectrum that extends to oceanographically unrealistic large wavelengths, i.e. a small bandwidth parameter $m_*$. The lack of horizontal structure and vertical velocity implies that the sum of an arbitrary number of such waves is a solution of the nonlinear equations of motion. The specification of an excessively small bandwidth $m_*$ ensures that the results for packet dispersion are scale invariant and thus leads to simple diagnostics. In this paper we employ a stratification ($N$) that is 4 times larger than the nominal 3 cph metric used in Polzin and Lvov (2017) in order to move the spectral boundaries away from the initial release site. The large scale background consists of a random inertial wavefield:

$$\mathcal{U}(z,t) = \sum_i U_i \, \sin(M_i z - ft + \phi_i) \,. \qquad (2)$$

with $1/40 \leq M_i/j_1 \leq 1600$ and the mode one equivalent $j_1 = \pi/1300$ m$^{-1}$. In contrast to Polzin and Lvov (2017) a slightly more sophisticated scale separation between wave and background (3) is enforced by a single pole filter:

$$U_i \to U_i \times \sqrt{M_i^2/((M_i^2 + (j_1/32)^2) * (1 + (ssM_i/m)^2))} \,, \qquad (3)$$

with time dependent $m$ and variable scale selection factor $ss$. We regard this as nothing more than an *ad hoc* device to enforce a spatial smoothing on the envelope scale, when information on the envelope has long since been discarded (see Section 3.2.1 of LP), in order to assess the sensitivity of the mean drift in wavenumber and dispersion of packets about that drift. Run parameters appear in Table 1.

This formulation has been selected with the intent that the vertical wavenumber gradient spectra are independent of vertical wavenumber for $j_1/32 \ll M_i \ll m/ss$. The results can be directly related to the GM76 spectrum, in which the level of the vertical wavenumber shear spectrum is proportional to the combination $e_0 m_*$, independent of vertical wavenumber $m$, where $e_0$ is the total energy (nominally 0.0030 m$^2$ s$^{-2}$) and $m_*$ is the vertical wavenumber bandwidth parameter (nominally $4\pi/1300$ m$^{-1}$) [1]. The ray tracing results will be presented as increments of $GM$, but what is intended is the dynamically relevant combination $e_0 m_*$. This two sided vertical wavenumber

---

[1] The one-sided vertical wavenumber - frequency energy spectrum is $e(m, \sigma) = e_0 \frac{2m_*}{\pi} \frac{1}{m_*^2 + m^2} \frac{2f}{\pi} \frac{1}{\sigma(\sigma^2 - f^2)^{1/2}}$ with $n(\mathbf{p}) = e(\mathbf{p})/\sigma_\mathbf{p}$.

Table 1. Run summary

We create an ensemble of test wave time series with #tw realizations of the randomly phased inertial field having #bw constituents uniformly distributed in vertical wavenumber between $j_{min} \times \pi/1300$ m and $j_{max} \times \pi/1300$ m. The length of the simulations is denoted by the number of time steps at the indicated time difference. The initial condition $m(t=0)$ is quoted in equivalent mode number $j = m\pi/1300$ m.

| name | $e_0/e_0^{GM}$ | time step | #tw | #bw | ss | $j_{min}$ | $j_{max}$ | $m(t=0)$ |
|---|---|---|---|---|---|---|---|---|
| run j3 | $10^0$ | 5000@1/10N | 20000 | 24001 | 1 | 0.025 | 1600 | $j = 100$ |
| run k3 | $10^0$ | 5000@1/10N | 20000 | 24001 | $1/\pi$ | 0.025 | 1600 | $j = 100$ |
| run l3 | $10^0$ | 5000@1/10N | 20000 | 24001 | $\pi$ | 0.025 | 1600 | $j = 100$ |
| run ll3 | $10^0$ | 5000@1/10N | 20000 | 24001 | $2\pi$ | 0.025 | 1600 | $j = 100$ |
| run m3 | 0.50 | 10000@1/10N | 20000 | 24001 | 1 | 0.025 | 1600 | $j = 100$ |
| run n3 | 0.50 | 10000@1/10N | 20000 | 24001 | $\pi$ | 0.025 | 1600 | $j = 100$ |
| run p3 | 0.25 | 20000@1/10N | 20000 | 24001 | 1 | 0.025 | 1600 | $j = 100$ |
| run q3 | 0.25 | 20000@1/10N | 20000 | 24001 | $\pi$ | 0.025 | 1600 | $j = 100$ |
| run g3 | $10^{-1}$ | 20000@1/10N | 10000 | 24001 | 1 | 0.025 | 1600 | $j = 100$ |
| run h3 | $10^{-1}$ | 20000@1/10N | 10000 | 24001 | $1/\pi$ | 0.025 | 1600 | $j = 100$ |
| run i3 | $10^{-1}$ | 20000@1/10N | 10000 | 24001 | $\pi$ | 0.025 | 1600 | $j = 100$ |
| run $a3_1$ | $10^{-2}$ | 240000@1/10N | 2000 | 24001 | 1 | 0.025 | 1600 | $j = 100$ |
| run $a3_2$ | $10^{-2}$ | 240000@1/10N | 2000 | 24001 | 1 | 0.025 | 1600 | $j = 100$ |
| run $a3_3$ | $10^{-2}$ | 240000@1/10N | 2000 | 24001 | 1 | 0.025 | 1600 | $j = 100$ |
| run $a3_4$ | $10^{-2}$ | 240000@1/10N | 2000 | 24001 | 1 | 0.025 | 1600 | $j = 100$ |
| run $a3_5$ | $10^{-2}$ | 240000@1/10N | 2000 | 24001 | 1 | 0.025 | 1600 | $j = 100$ |
| run $d3_1$ | $10^{-3}$ | 800000@1/5N | 1000 | 24001 | 1 | 0.025 | 1600 | $j = 100$ |
| run $d3_2$ | $10^{-3}$ | 800000@1/5N | 1000 | 24001 | 1 | 0.025 | 1600 | $j = 100$ |
| run $d3_3$ | $10^{-3}$ | 800000@1/5N | 1000 | 24001 | 1 | 0.025 | 1600 | $j = 100$ |
| run $d3_4$ | $10^{-3}$ | 800000@1/5N | 1000 | 24001 | 1 | 0.025 | 1600 | $j = 100$ |
| run $d3_5$ | $10^{-3}$ | 800000@1/5N | 1000 | 24001 | 1 | 0.025 | 1600 | $j = 100$ |
| run $d3_6$ | $10^{-3}$ | 800000@1/5N | 1000 | 24001 | 1 | 0.025 | 1600 | $j = 100$ |
| run $d3_7$ | $10^{-3}$ | 800000@1/5N | 1000 | 24001 | 1 | 0.025 | 1600 | $j = 100$ |
| run $d3_8$ | $10^{-3}$ | 800000@1/5N | 1000 | 24001 | 1 | 0.025 | 1600 | $j = 100$ |
| run $d3_9$ | $10^{-3}$ | 800000@1/5N | 1000 | 24001 | 1 | 0.025 | 1600 | $j = 100$ |
| run $d3_{10}$ | $10^{-3}$ | 800000@1/5N | 1000 | 24001 | 1 | 0.025 | 1600 | $j = 100$ |
| run e3 | $10^{-3}$ | 800000@1/5N | 1000 | 24001 | $1/\pi$ | 0.025 | 1600 | $j = 100$ |
| run f3 | $10^{-3}$ | 800000@1/5N | 1000 | 24001 | $\pi$ | 0.025 | 1600 | $j = 100$ |

Power Spectral Density (PSD) of one horizontal component shear has an asymptotic level of:

$$PSD(\text{GM, two-sided, one component shear}) = \frac{3}{4\pi} m_* e_0 \,. \qquad (4)$$

At oceanic levels the one-sided vertical wavenumber power spectrum of two-component shear is approximately $1.0 N_0^2$ m$^{-1}$ with $N_0 = 3$ cph.

Test waves are traced in $m$ (figure 1) and in $z$ (figure 2) as a function of time using a simple forward difference scheme in (1). Values of the random phase $\phi_i$ are stored to permit quantification of the total phase $M_i z - ft + \phi_i$ in (2). Ensemble averages are generated by averaging over tens of thousands of test waves.

Test waves are released at a vertical wavenumber $m$ equivalent to mode-50 ($50\pi/1300$ m$^{-1}$). Analysis starts at a time $t = 0$ when the wave packet crosses the equivalent mode-100, $m_0 \equiv m(t=0) = 100\pi/1300$. The analysis period ends when a small fraction ($< 1\%$) of the wave packets have intrinsic frequencies larger than $N/\sqrt{2}$ or intrinsic frequency smaller than $\sqrt{2}f$, figure 1. These metrics signify departures of the dispersion relation from its non-rotating hydrostatic approximation and thus the absence of scale invariant behavior. Despite such conditioning, subtle non-scale invariant behavior is noted in the latter half of all simulations.

## 3. Metrics of Transport

*a. Kinetic Equation*

1) FOKKER-PLANK EQUATION

The moment method is a methodology for the interpretation of a Fokker-Planck equation. For internal waves this is eq. 2.20 of LP. For our one-dimensional model, equation 2.20 of LP reduces to:

$$\frac{\partial}{\partial t} n(m,t) + \frac{\partial}{\partial m} D_{33}(m) \frac{\partial}{\partial m} n(m,t) = 0 . \tag{5}$$

The vertical-vertical component of the diffusivity tensor $D_{33}$ (eq. 47; Polzin and Lvov (2017)) from the kinetic equation is

$$D_{33} = \pi k^2 f \int d\mathbf{p}_1 n(\mathbf{p}_1) m_1^2 \delta(m_1 \frac{\sigma}{m} - \sigma_1) . \tag{6}$$

The action density $n(\mathbf{p})$ represents the high-frequency field, $n(\mathbf{p}_1)$ the inertial background. Our inertial wave model is two dimensional in the $x - z$ plane, so that $n(\mathbf{p}_1) = n(k_1, 0, m_1) \delta(l_1)$ in which $n(k_1, m_1) = \frac{1}{4} \frac{e(\sigma_1, m_1)}{\sigma_1} \frac{d\sigma_1}{dk_1}$, and the corresponding normalized frequency spectrum is $\delta(\sigma_1 - f)$. The diffusivity is estimated by integrating over horizontal azimuth, changing variables from horizontal wavenumber magnitude to wave frequency and integrating over vertical wavenumber. The factor $e_0$ represents the total internal wave energy, kinetic plus potential. These are in a ratio of 3:1 for the GM76 model. Our inertial wave model has no potential energy. Incorporating this into (6), we find

$$D_{33}^{1D} = \frac{3}{8} \frac{km^2 e_0 m_*}{N} . \tag{7}$$

2) Moments

The moment method proceeds by multiplying the diffusion equation by $m^j$, utilizing the chain rule, integrating over the spectral domain and discarding terms at $m = \pm\infty$ to produce differential equations for the $j^{th}$ moment. Here $\langle\ldots\rangle$ represents the integral over vertical wavenumber.

$$\langle n(m) \rangle_t = 0 , \tag{8}$$

$$\langle mn(m) \rangle_t = \langle \partial_m (D_{33}) n(m) \rangle , \tag{9}$$

$$\langle m^2 n(m) \rangle_t = \langle \partial_m (2m D_{33}) n(m) \rangle . \tag{10}$$

These moments have analytic solutions for our scale invariant model:

$$\langle m \rangle = m_0 \, e^{2D_{33}t/m^2} , \tag{11}$$

$$\langle m^2 \rangle = m_0^2 \, e^{6D_{33}t/m^2} , \tag{12}$$

$$\langle (m - \langle m \rangle)^2 \rangle = \langle m^2 \rangle - \langle m \rangle^2 . \tag{13}$$

*b. Ray Path Methods*

1) Fokker-Plank Equation

In LP we use ray path techniques to formulate an ensemble average transport equation:

$$\frac{\partial \langle n_{\mathbf{p}} \rangle}{\partial t} = -\nabla_{\mathbf{p}_i} \cdot C_{ij}(\mathbf{p}) \cdot \nabla_{\mathbf{p}_j} \langle n_{\mathbf{p}} \rangle - \nabla_{\mathbf{p}_i} \langle \dot{\mathbf{p}}_i \rangle \langle n_{\mathbf{p}} \rangle. \tag{14}$$

with time integrated lagged auto-correlation function

$$C_{ij}(\mathbf{p}) = \int_{t-\tau}^{t} dt' \left\langle \left[ \dot{\mathbf{p}}(\mathbf{r}(t)) - \langle \dot{\mathbf{p}}(\mathbf{r}(t)) \rangle \right]_i \left[ \dot{\mathbf{p}}(\mathbf{r}(t')) - \langle \dot{\mathbf{p}}(\mathbf{r}(t')) \rangle \right]_j \right\rangle. \tag{15}$$

Note that both first and second moments of the ensemble average action density appear explicitly in the transport equation.

## 2) A Path Integral for the Mean Drift

Here we present a derivation for the mean drift $\langle \dot{m} \rangle$. We start at (1) and represent the background shear with its inverse Fourier transform:

$$\dot{m} = -k\mathcal{U}_z = -\frac{k}{2\pi} \int_{-\infty}^{\infty} \Upsilon_z(M) e^{i(Mz-ft)} dM . \tag{16}$$

The factor $\Upsilon_z$ represents the Fourier coefficient for vertical shear (4). The vertical coordinate following the ray path is the time integral of the vertical group velocity, $C_g^z$. For internal waves:

$$z(t) = \int_{-\infty}^{t} \frac{-kN}{m(t')^2} dt' .$$

The major contributions to the integral (16) come from conditions in which the inertial phase velocity $f/M$ is equal to the internal wave group velocity, $C_g^z = -sgn(m)|k|N/m^2$. In the ray coordinate, $z$ is a function of time and integration in time permits application of a stationary phase approximation. There is a phase:

$$\vartheta(t) = -M \int_{-\infty}^{t} dt' \frac{kN}{m^2} - ft . \tag{17}$$

Differentiating with respect to time:

$$\dot{\vartheta} = -M\frac{kN}{m^2} - f \tag{18}$$

and differentiating once more:

$$\ddot{\vartheta} = M\frac{2kN}{m^3} \dot{m} . \tag{19}$$

After Taylor series expanding the phase about the resonance $M_r = -fm^2/kN$,

$$e^{i\vartheta} \to e^{i[\vartheta(t_0) + \ddot{\vartheta}(t_0)(t-t_0)^2/2 + \ldots]} .$$

Changing the variable of integration from $M$ to $t$ by considering the background wavenumber $M$ to be a property of the time evolving resonance along the ray path returns,

$$\int dM \to \int dM_r \Big|_{t=t_0} + \int \frac{dM_r}{dt}\Big|_{t=t_0} dt \ldots$$

and applying the stationary phase formula (Bender and Orzag (1978)), we obtain

$$1 \cong -k\sqrt{\frac{\pi m^3}{kN\dot{m}}} \frac{2mf}{kN} \left[\frac{1}{2\pi} \frac{\|\Upsilon_z(M_r)\|}{\sqrt{M_r}}\right] \cos(\pm\frac{\pi}{4} + \vartheta(t_0)) , \qquad (20)$$

with choice of sign depending upon the value of $\ddot{\vartheta}(t_0)$. We square both sides and average over a vertical wavelength,

$$\langle \rangle = \frac{1}{\lambda_v} \int_{-\lambda_v/2}^{+\lambda_v/2} dz = \frac{1}{\lambda_v} \int_{-\pi}^{+\pi} \frac{dz}{d\vartheta} d\vartheta .$$

This average represents a sum over all possible paths encoded in the phase function $\vartheta$ and is a 'path integral'. Upon recognizing the definition of the Power Spectral Density (PSD) given the choice of $\lambda_v$ for a transform interval,

$$\frac{2\pi}{\lambda_v}[(\frac{1}{2\pi})^2 \Upsilon_z(M_r)\Upsilon_z^*(M_r)] \equiv PSD ,$$

we obtain, for the GM model in which $PSD = \frac{3}{4\pi}m_*e_0$,

$$\langle \dot{m} \rangle = \frac{3}{4}\frac{kmm_*e_0}{N} .$$

The mean drift is equivalent to $\partial_m D_{33}^{1D}$ derived from kinetic theory (7). Note that the mean drift is arrived at as a resonant process.

3) Lag-Correlation Functions and Second Moments

The lagged auto-correlation function:

$$\frac{1}{2}\frac{d(m-\langle m\rangle)^2}{dt} = k^2 \int_{t-\tau}^{t} (\mathcal{U}_z(t) - \langle \mathcal{U}_z(t)\rangle)(\mathcal{U}_z(t-\tau) - \langle \mathcal{U}_z(t-\tau)\rangle)d\tau \tag{21}$$

is the one-dimensional representation for $C_{33}$ (15). A prediction for the lag-correlation function and thus variance can be obtained by noting that the highest frequency encounters are associated with the smallest scale background waves and that these are essentially stationary in the time required for a high-frequency wave packet to propagate through them. This suggests a 'frozen field' hypothesis in which the encounter frequency $s$ is identified as:

$$s = MC_g^z, \tag{22}$$

where $C_g^z$ is vertical group velocity. We investigate by assuming ergotic statistics at small amplitude and invoking the Wiener-Kintchen theorem. Thus

$$\frac{1}{2}\frac{d(m-\langle m\rangle)^2}{dt} = k^2 \int_{t-\tau}^{t}\int_0^{+\infty} \cos(s\tau)P_{\mathcal{U}_z}(s)ds d\tau, \tag{23}$$

in which $s$ is the encounter frequency along the ray and $P_{\mathcal{U}_z}$ is the shear spectral density in that coordinate. From a purely empirical standpoint, the spectra of vertical shear in encounter frequency $s$ (figure 3) are bandwidth-limited and white. Much of the variability collapses onto:

$$P_{\mathcal{U}_z}(s) = \frac{3}{4\pi}\frac{e_0 m_*/C_g^z}{1+(\text{ss}\frac{s}{\langle\omega\rangle})^2}. \tag{24}$$

which can be arrived at by using (3) and the change of variables (22). Carrying out the cosine transform, we obtain:

$$\frac{1}{2}\frac{d(m-\langle m\rangle)^2}{dt} \cong k^2 \frac{3}{8}\frac{e_0 m_*}{C_g^z}\frac{\langle\omega\rangle}{\text{ss}}\int_{t-\tau}^{t} e^{-\langle\omega\rangle t/\text{ss}} d\tau. \tag{25}$$

If $\langle\omega\rangle$ is regarded as a constant,

$$\frac{1}{2}\frac{d(m-\langle m\rangle)^2}{dt} \cong \frac{3}{4}\frac{km^2 e_0 m_*}{N}[1 - e^{\langle\omega\rangle t/\text{ss}}]. \tag{26}$$

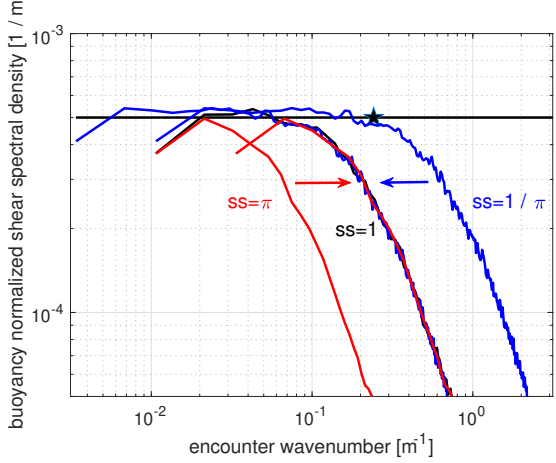

Figure 3. Example of a frequency spectrum of vertical shear following a ray path for backgrounds with $GM \times 10^{-3}$, converted to an encounter wavenumber using the group velocity. Results for scale separation factors $ss = [\pi, 1, \pi^{-1}]$ are indicated using red, black and blue traces. The spectra are collapsed after re-scaling the frequency by $ss$ (indicated by arrows) and the black trace is buried. Such spectra are band-width limited and white with a 1/2 power point at $m(t=0)$ for $ss = 1$ denoted by the pentagram. The black horizontal line represents one-side of the GM vertical wavenumber shear spectrum, $0.001 e_0 m_* N^{-2} = 0.5 \times 10^{-3}$ / m$^{-1}$. This spectral description is essentially consistent with using the time domain Fourier transform pair of an exponential auto-correlation function.

This prompts several interpretations. The first is the identification of a diffusivity. In the limit that $\langle m \rangle / \langle \dot{m} \rangle \gg ss/\langle \omega \rangle \equiv \tau_c$, the diffusivity is just

$$D_{33} \cong k^2 \langle \mathcal{U}_z^2 \rangle \tau_c \quad \text{with} \quad \tau_c = \frac{ss}{\langle \omega \rangle} \quad \text{and} \quad \langle \mathcal{U}_z^2 \rangle = \frac{3}{4\pi} e_0 m_* \int_0^\infty \frac{dM}{1 + (ss\frac{M_i}{m})^2} = \frac{3}{8} \frac{m}{ss} e_0 m_* ,$$

which is equivalent to that arrived at from resonant theory (7). Note that $\tau_c$ is controlled by the scale separation criterion $ss$. However, since the encounter spectrum of vertical shear is white, it really doesn't matter what the scale selection factor $ss$ is, the product $\langle \mathcal{U}_z^2 \rangle \tau_c$ is a constant for a white spectrum, as long as the integral (25) has converged. Conversely, if the background vertical shear spectrum was other than uniform in vertical wavenumber, variations in the scale separation $ss$ would have a weak influence on the diffusivity.

The second interpretation is that the long time limit in (25), in which effects of the mean drift on the ensemble average frequency $\langle \omega \rangle$ are neglected, is identifiable as the Markov approximation. Our numerical results suggest that this approximation is challenged at oceanic amplitudes.

The third interpretation is that the diffusivity results from a non-resonant process: the bandwidth of the integrand in (25) is not the resonant bandwidth from the ray numerics in figure 5 or from kinetic theory. This result is in direct contrast to McComas and Bretherton (1977); Nazarenko et al. (2001), both of whom present derivations that reduce the diffusivity to a delta function representation of the resonant manifold. Those derivations implicitly assume the background is statistically homogeneous in the spectral domain, omit an explicit reference to an ensemble averaging process and associated time dependent ensemble average drift $\langle \dot{m} \rangle$ in (21). They then utilize a Taylor series expansion about a *time invariant resonance* after representing the vertical wavenumber tendency $\dot{m}$ in terms of its inverse Fourier transform (16).

## 4. Numerical Results

Our diagnostics include numerical evaluations of probability distributions of inertial wave phase sampled by the high-frequency waves, moments of test wave vertical wavenumber and lag-correlation analyses in test wave vertical wavenumber. The probability distributions demonstrate phase locking about the resonant phase velocity equals group velocity condition even at oceanic amplitudes. The moment analysis quantifies an inhibition of the second moment at oceanic amplitudes. The first moment, upon which the ray tracing closure (14) hinges, is less sensitive, but departures are still noted. The lag-correlation analysis quantifies the departure of the second moment from its resonant prediction as being a competition between the mean drift and the correlation time scale imposed by an *ad hoc* scale separation criterion (3).

*a. Phase Locking*

1) PHASE PROBABILITY DISTRIBUTIONS

Kinetic equations assume a zeroth order description in which wave phases are uncorrelated and then predict action transfer associated with phase locking at first order. The inference of phase locking is indirect as one is closing out a hierarchy of moments.

In ray tracing, phase locking can be much more directly assessed (Fig. 4). In this one-dimensional system, the probability density $\hat{p}$ of background phase $\vartheta_i$ (2) is estimated as a function of background wavenumber $M_i$:

$$\vartheta_i = M_i z - f t + \phi_i \, ,$$

when the test wave has the value $m_0 - \delta \leq m \leq m_0 + \delta$,

$$\hat{p} = p[(\vartheta, M) \mid m = m_0 \pm \delta] \, ,$$

with $m_0$ equal to the equivalent of mode-100. Since the background shear is specified as a sum of cosines, a probability density maximum centered on either $0-\pi$ or $\pi-2\pi$ implies a bound wave behavior in which the test wave preferentially occupies a background crest or trough and oscillates about the crest/trough. Probability extrema centered on $\pi/2-3\pi/2$ implies a non-zero average shear and drift to either larger or small scales. In all our runs the probability extrema occur in association with the resonance $MC_g^z = f$ and maxima are located about $\pi/2 < \vartheta < 3\pi/2$, indicating a net drift of test waves to smaller scales, figure 4. The shoulders of the resonance appear more representative of a bound wave behavior.

2) BANDWIDTH AND MEAN SHEAR

A quantitative measure of the downscale transport ensemble average $\langle \dot{m} \rangle = -k\langle \mathcal{U}_z \rangle$ can be obtained from the phase distributions in figure 4. The ensemble averaged shear is

$$\langle \mathcal{U}_z \rangle \propto \int_{-\infty}^{+\infty} dM \int_0^{2\pi} \hat{p}(\vartheta, M) \cos(\vartheta) d\vartheta , \qquad (27)$$

so that the phase average,

$$\int_0^{2\pi} \hat{p}(\vartheta, M) \cos(\vartheta) d\vartheta , \qquad (28)$$

provides a metric of the amplitude and bandwidth of the energy transfer process. These phase averaged distributions (figure 5) are neither peaked precisely at the nominal resonance nor are the distributions symmetric about the peak. However, the half-widths $\gamma_M$ are reasonably well predicted by (Polzin and Lvov 2017):

$$\frac{\gamma_M}{M_r} \cong [3\pi(\frac{\omega}{Nf})^2 e_0 m_* M_r]^{1/3} . \qquad (29)$$

This scaling of the resonant well differs from the small amplitude limit of the kinetic equation, $\gamma \propto e_o/e_o^{GM}$ and the finite amplitude degradation into the rms Doppler shift, $\gamma \propto (e_o/e_o^{GM})^{1/2}$, (Polzin and Lvov 2017). These differences in scaling underscores the fact that ray tracing is a frequency-modulated (FM) paradigm and the kinetic equation represents nonlinear transfers as an amplitude-modulated (AM) process.

Estimates of the mean shear can be obtained by integrating (27) over phase $\vartheta$, figure 5, and subsequently integrating over background wavenumber $M$, figure 6. We anticipate a scaling for the downscale energy transport $\mathcal{P}$ (Section 5.a) in which $\mathcal{P} = 2\int_f^N \langle \dot{m} \rangle e(m, \sigma) d\sigma \propto e_0^2$. Since

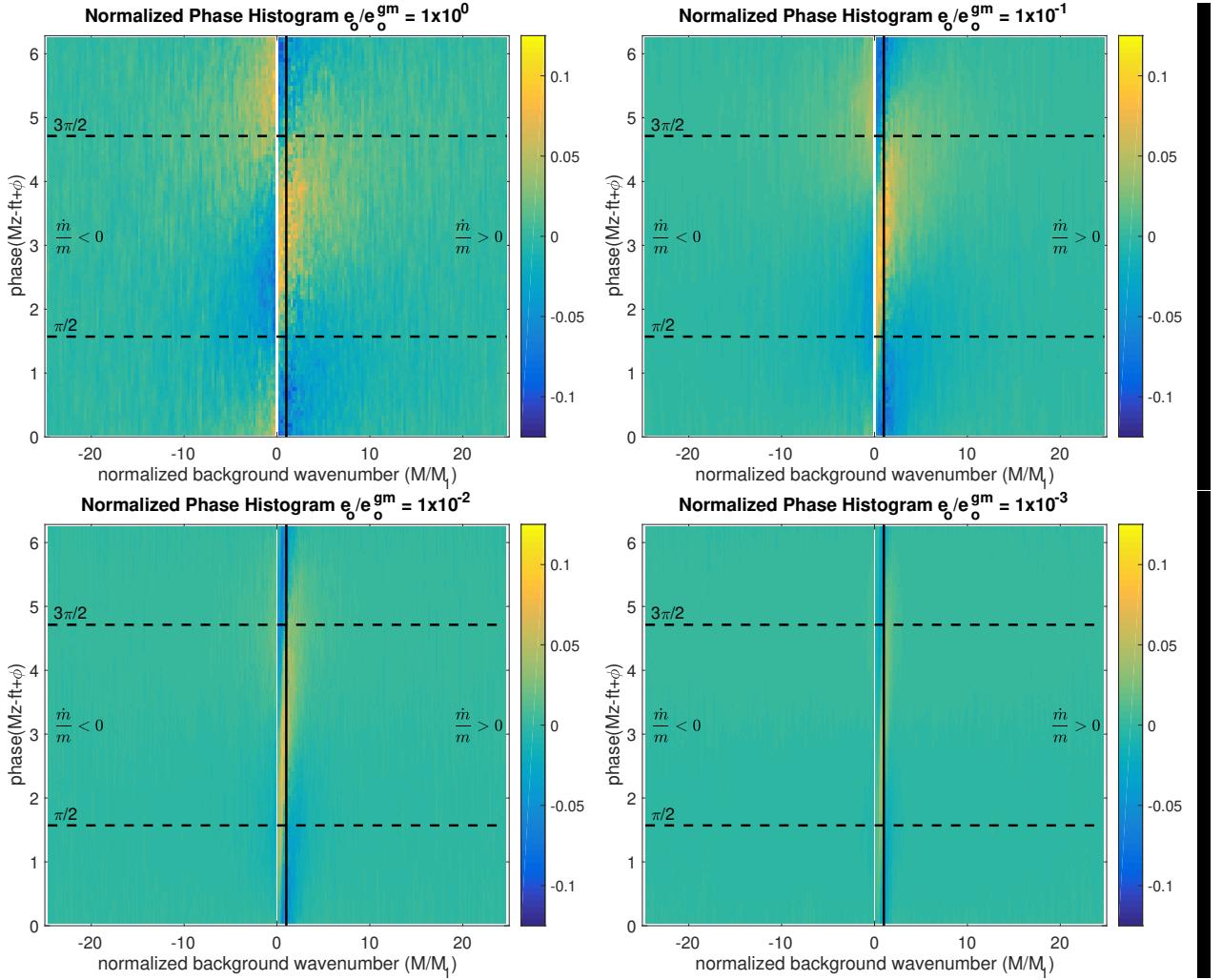

Figure 4. Deviations of the phase $\vartheta_i = M_i z - ft + \phi_i$ from a uniform distribution as a function of background wavenumber $M_i$ normalized by resonance condition $f/C_g$. The histograms are normalized such that a value of 0.1 indicates a 10% increase in the probability density. The vertical black lines denote the approximate resonance condition $C_{ph} = C_g^z$. Upper left (upper right, lower left, lower right) panels are $10^0 GM$ ($10^{-1} GM$, $10^{-2} GM$, $10^{-3} GM$).

$e(m, \sigma) \propto e_0$ and the root-mean-square inertial shear $\mathcal{U}_z^{\text{rms}}$ scales as $e_0^{1/2}$, we anticipate $\langle \mathcal{U}_z \rangle / \mathcal{U}_z^{\text{rms}}$ to scale as $e_0^{1/2}$. Starting at a level of 0.1 GM, we find a factor of two departure from this scaling at oceanic amplitudes, figure 6, with larger scale separation factors *ss* associated with greater departures from the scaling. This factor of two decrease is insufficient to overcome the order of magnitude disparity between predicted transports (Section 5.a) and observations.

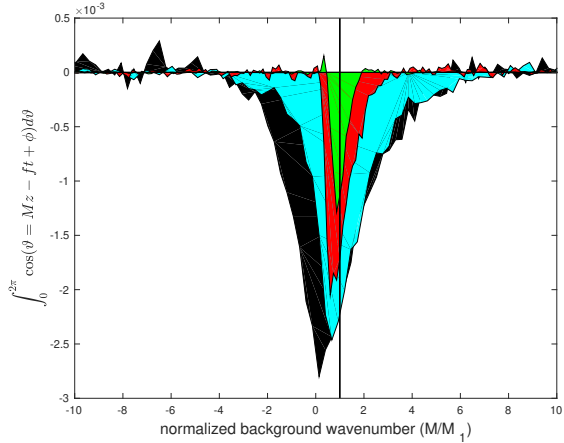

Figure 5. Phase averaged energy transfer distributions (28), $e_o/e_{GM} = [1, 0.1, 0.01, 0.001]$ [black, cyan, red, green]. The widths of the distributions varies more than the peak height, consistent with the scaling described in Section a. The vertical line represents the resonance $M_r = f/C_g^z$ associated with the group velocity $C_g^z$ at time $t = 0$.

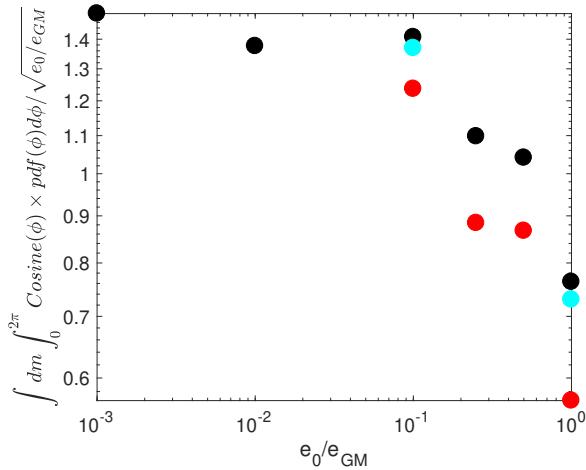

Figure 6. Normalized version of the mean drift rates (27) estimated from the phase distributions in figure 4, as a function of $e_o/e_{GM}$. The normalized values would appear as a constant if $\langle \dot{m} \rangle \propto e_o$. Scale separations $ss = [1, 1/\pi, \pi]$ are visualized in [black, cyan, red]. Greater scale separations imply larger correlation time scales, hence smaller differences between mean drift and correlation time scales, and thus larger departures from the resonant scaling of the mean drift (2).

Contributions to the mean shear (27) can come from either an increasing bandwidth $\gamma_M$ and/or increasing peak probability density. Given that the bandwidth of the probability distribution (29) scales as $e_0^{1/3}$, implied in the scaling $\langle u_z \rangle / u_z^{\rm rms} \propto e_0^{1/2}$ is that the peak amplitude of the phase averaged distributions scale as $e_0^{1/6}$. Indeed, we find a factor of two increase in peak probability density as the background increases from 0.001 GM to 0.1 GM, figure 5.

*b. Moments and Correlation Functions*

In this sub-section we quantify lag-correlation functions (figure 7) and the evolving moments in vertical wavenumber of a test wave distribution (figure 8). There are contrasts between the wave turbulence and ray tracing paradigms. From a wave turbulence perspective, everything is subsumed into the resonant process. From a ray path perspective, the correlation time scale results from the non-resonant parts of the problem whereas the mean drift is related to zero encounter frequency in which inertial phase velocity equals internal wave group velocity.

The lag-correlation functions (figure 7) are suggestive of a two time scale process, even at small amplitude. There is rapid evolution of the time integrated correlation functions on a short time scale of approximately $1/\langle \omega \rangle$ and a slower evolution on a longer time scale. Note that, although correlations are small at large lag, integration demonstrates these large lags have non-zero contributions. We identify the fast time scale process as the non-resonant response and dispersion about the mean drift. This fast time scale is controlled by the scale separation factor *ss* (25). We intuit the slow time scale to be associated with the resonant response and mean drift. Such contributions at large lags are robust in the sense that we have subtracted the sample mean, consistent with (21). Rescaling $\mathcal{U}_z(t)$ to account for non-stationary statistics does not eliminate the increasing trend in the time integrated correlation. A possible metric of this time scale is the width of the resonant well, (29), divided by the mean drift $\langle \dot{m} \rangle$ projected onto the resonant wavenumber $M_r = fm/\omega$. We are unable to further elucidate this long time scale as our simulations terminate before it is resolved.

Deviations from the nominal scalings (13) for both first and second moments are discernible in all model runs (figure 8, compare green and red traces). Such departures are much more subtle at small amplitude and further diagnostics demonstrate that these departures occur in the latter half of the simulations. We suspect deviations from this scaling at small amplitude are related to the use of $(\omega^2 - f^2)^{1/2} = kN/m$ as a dispersion relation associated with the ray equations (1) rather than the scale invariant $\omega = kN/m$. We have terminated the analysis such that less than 1% of wavves

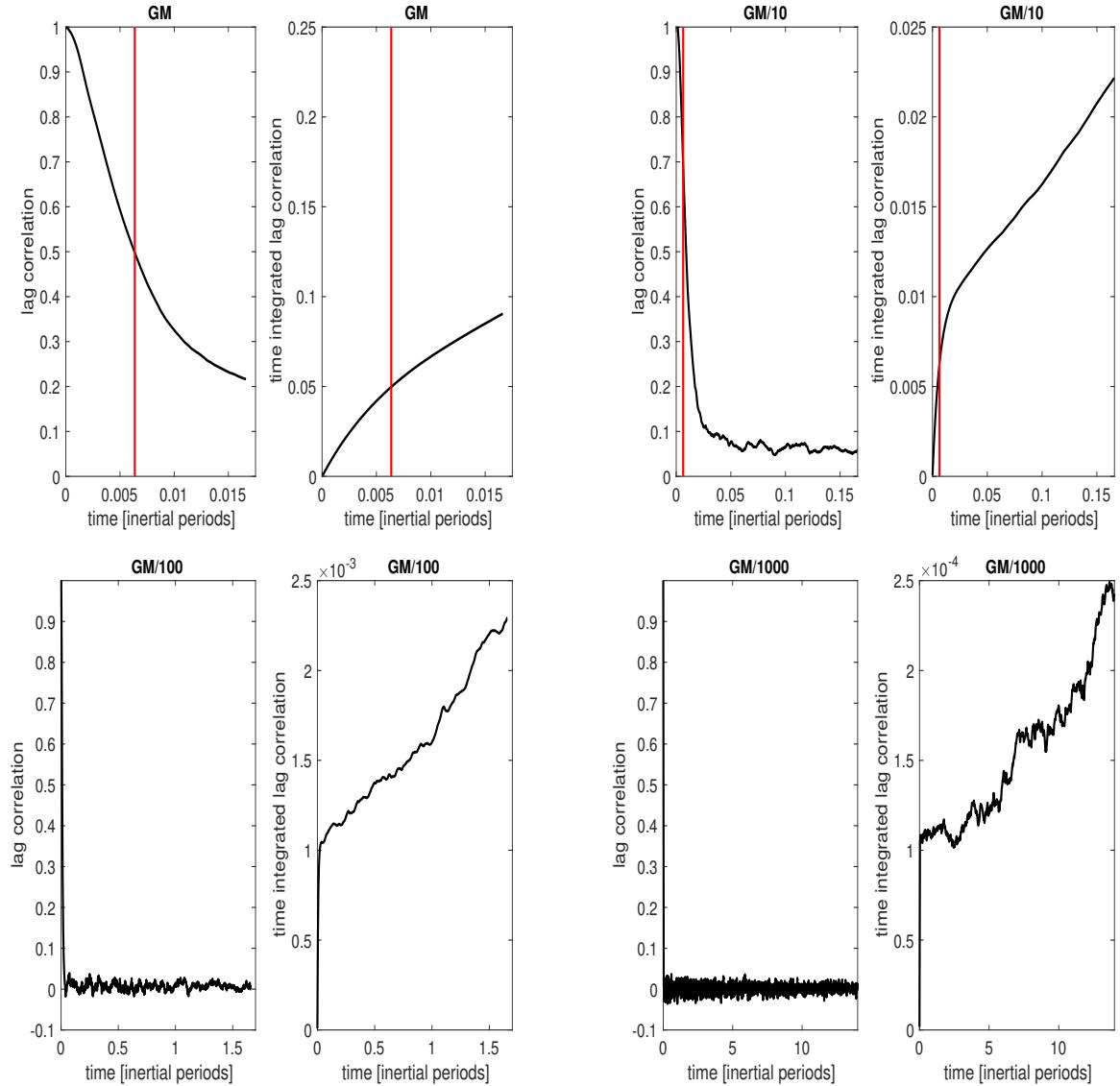

Figure 7. Lagged autocorrelation functions (left panels) and their time integrals (right panels). Upper left (upper right, lower left, lower right) panels are $10^0 GM$ ($10^{-1} GM$, $10^{-2} GM$, $10^{-3} GM$). The red vertical line represents the wave frequency $\omega$ at $t = 0$, which, in turn, represents the correlation time scale. Time integration reveals tertiary contributions at times much larger than the non-resonant correlation time scale. Our hypothesis is that this tertiary contribution comes from the time changing structure of the resonant well.

have $\omega < \sqrt{2} f$, but this might not be sufficiently stringent. The deviations at oceanic amplitudes are of greater import. At oceanic amplitudes the evolution rates of wavenumber and frequency are of similar order of magnitude to the wave frequency, with the consequence that the long time approximation in (25) is no longer accurate. That is, we find an inhibition of the second moment when the drift time scale $m/\langle \dot{m} \rangle$ is similar to the correlation time scale. Consistently, larger scale separations $ss$ lead to earlier onset of departures from the nominal scaling at smaller background amplitudes, figure (6). The long time limit in which $\langle \omega \rangle$ can be considered to be constant in (25) is known as the Markov approximation and the suppression of first and second moments occurs in conjunction with the transition to a non-Markovian limit. That this should impact the second moment is obvious from (25), but the rational for departures in the mean drift is less obvious. An explanation likely lies in higher order contributions to the stationary phase analysis in Section 2

*c. Reprise*

In this section we have presented quantitative diagnostics from one-dimensional ray tracing simulations. This presentation documents that ray theory brings results that differ from the kinetic equation. Scalings for the resonant bandwidth are different, ray tracing has a mean drift absent from the kinetic equation, and dispersion about that mean drift results from a non-resonant process in ray theory. Similar behavior should be recoverable from 'kitchen sink' treatments (Henyey et al. 1986; Sun and Kunze 1999b; Ijichi and Hibiya 2017) if those numerics were executed in an appropriate small amplitude parameter regime.

## 5. Discussion

In this section we compare a prediction for the rate at which energy is supplied to internal wave breaking processes with an observational metric. The prediction is an order of magnitude larger than the observations. We then discuss this overprediction in the context of 'kitchen sink' treatments of ray tracing and reflect on the overprediction in the context of extreme scale-separated physics discarded by the ray tracing approximation.

*a. Energy Transport*

In the Introduction we noted tension between an apparent pattern match between observed spectral power laws being in apparent agreement with stationary states of the Fokker-Planck equation (Polzin and Lvov 2011) and this result being inconsistent with what is observationally understood about the energy sources and sinks (Polzin and Lvov 2017; Dematteis et al. 2022), in particular the status of the Garrett and Munk model (GM76) being a no-flux stationary state due to there there being no gradients in action in vertical wavenumber. The ray path perspective moves away from this interpretation so that downscale transport is closed as an advective transport (14). We evaulate (14) by equating the mean drift $\langle \dot{m} \rangle$ with the gradient of the vertical-vertical component of the diffusivity tensor, $\partial D_{33}/\partial m$.

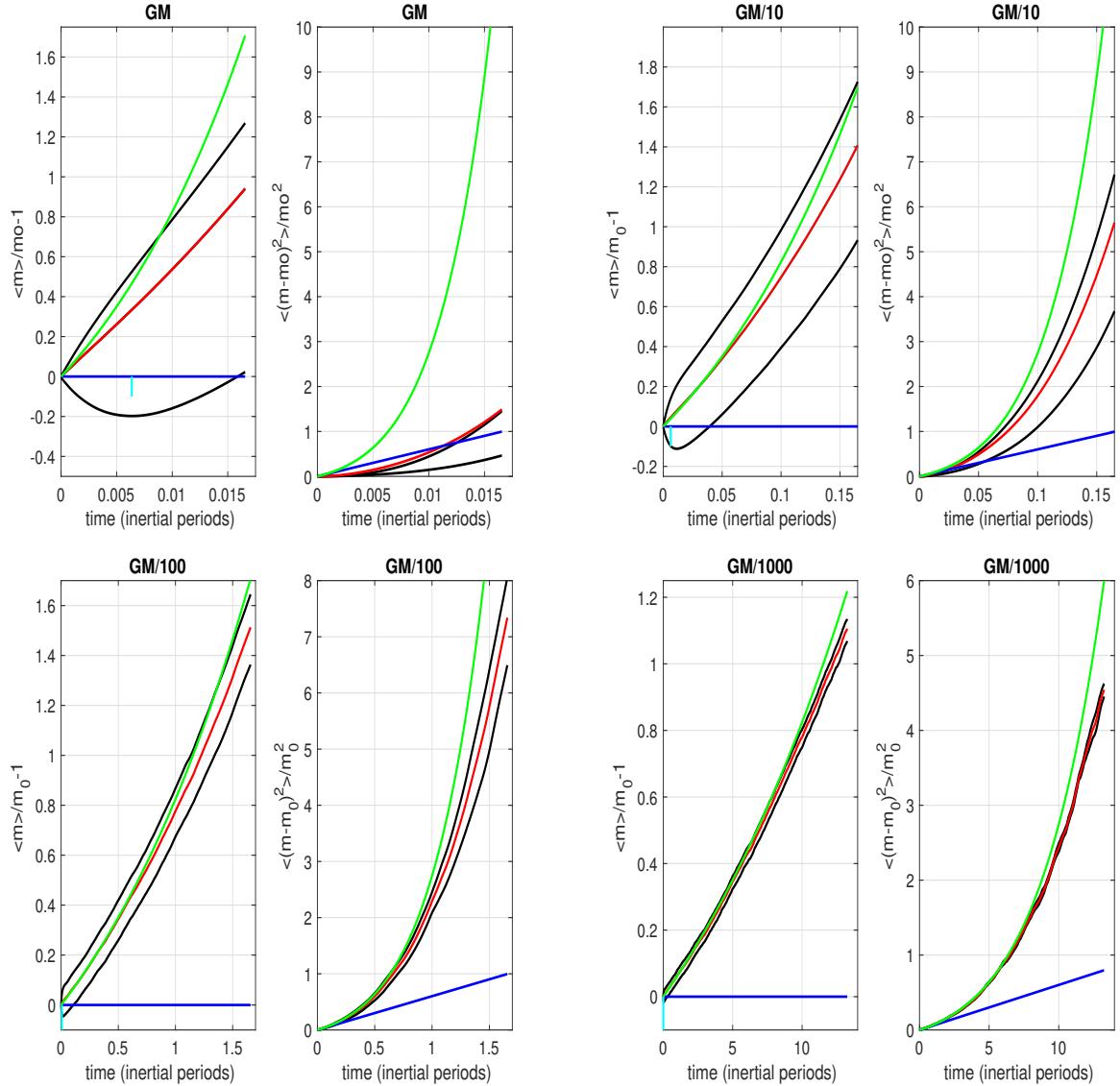

Figure 8. First and second moments vs time (red and black) with prediction based upon the Eulerian paradigm (13) (green). The black lines are ensemble averages conditioned on the sign of the wavenumber tendency (velocity in wavenumber space) at $t = 0$. The red line represents the ensemble average of the conditional moments. The difference between the conditioned estimates relates to the lag-correlation time scale (25). The blue line represents a constant diffusivity model. Upper left (upper right, lower left, lower right) quadrants are $10^0$ ($10^{-1}$ ($10^{-2}$, $10^{-3}$) times GM. Left hand panels are the first moment. Right hand panels are the second moment. The cyan vertical line represents the inverse wave frequency at $t = 0$.

After identifying $n_{\mathbf{p}_1}$ in (6) with the GM76 spectrum, integrating over horizontal azimuth, changing variables from horizontal wavenumber magnitude to wave frequency and integrating over vertical wavenumber, in the limit that $m \gg m_*$ and $\sigma \gg f$, (6) becomes

$$D_{33}^{GM} = k^2 f e_0 m_* \frac{m_1^2}{m_*^2 + m_1^2} \frac{m}{\sigma} \int_f^\sigma \frac{2f}{\pi} \frac{d\sigma_1}{\sigma_1^2 \sqrt{\sigma_1^2 - f^2}} \to \frac{2}{\pi} \frac{k m^2 e_0 m_*}{N} \, , \qquad (30)$$

to be compared with (7). Here, as in (7), $k$ is horizontal wavenumber magnitude, $e_0$ is the total energy, $m_*$ is a bandwidth parameter and $N$ buoyancy frequency. In the GM models, frequency and vertical wavenumber energy spectra are regarded as separable and normalization constants are incorporated. Thus in (30), we have $\frac{2f}{\pi} \int_f^N \frac{d\sigma}{\sigma \sqrt{\sigma^2 - f^2}} \cong 1$. After including a factor of two to account for the two-sided spectral representation, the downscale energy transport is

$$\mathcal{P} = 2 \int_f^N \langle \dot{m} \rangle e(m, \sigma) d\sigma = \frac{8}{\pi} \left(\frac{2}{\pi}\right)^2 \left(\frac{e_0 m_*}{N}\right)^2 f \log\left(\frac{N}{f}\right) \cong 1.0 \times 10^{-8} [\text{W kg}^{-1}] \, . \qquad (31)$$

which, apart from the prefactor of $1.0 \times 10^{-8}$ being an order of magnitude too large, is virtually identical to the finescale parameterization (Polzin et al. 2014), their equations 27 and 40.

We believe that this one-dimensional treatment is reasonable representation of high-frequency wave refraction in near-inertial shear. The one-dimensional version dates to the dawn of modern oceanography and is supported by basic scale analysis (McComas and Bretherton 1977; Sun and Kunze 1999a). It is underpinned by the integrable singularity of the inertial peak in the internal wave frequency spectrum and the lack of horizontal velocity gradients in that peak that is encoded in the dispersion relation. Assessments of extreme scale-separated interactions in a non-rotating approximation (Dematteis et al. 2022) assigns diffusive transports associated with horizontal and off-diagonal components of the diffusivity tensor that are two orders of magnitude smaller than this advective transport associated with the vertical-vertical component. The non-rotating analysis over-estimates the importance of horizontal and off-diagonal transports in a rotating system due to the relative lack of horizontal velocity gradients in the inertial peak. Our model is idealized, but not unrealistic.

*b. Kitchen Sink Numerics*

Ray methods have also been used in a 'kitchen sink' manner in which test waves are traced in a background consistent with a spectrally filtered version of the Garrett and Munk frequency - vertical wavenumber spectrum (Henyey et al. (1986); Sun and Kunze (1999b); Ijichi and Hibiya (2017)). These studies regard the scale separation as a tunable parameter to arrive at advective estimates of downscale transport that, unlike our one-dimensional model, are in sensible agreement with the Finescale Parameterization. We offer two insights.

The first is that ray tracing is an asymptotic method requiring a scale separation in horizontal wavenumber in addition to the spatial averaging implied in the envelope structure of a wavepacket and a small amplitude assumption (LP). These kitchen sink numerical assessments consistently document a sensitivity to the specification of the scale separation and consistently find that the observed finescale metric of energy sourced to turbulent dissipation (Polzin et al. (2014)) requires a scale *equivalence*, i.e. requires the small parameter of an asymptotic expansion to be $\sim O(1)$. This is the hallmark of interactions that are spectrally local in wavenumber and best treated by other methods such as in Dematteis and Lvov (2021); Dematteis et al. (2022).

The second insight is that (31) and associated scaling is a fundamental metric that should be recoverable by kitchen sink efforts as a small amplitude limit using a scale separation that aligns with the assumptions underpinning ray tracing. Departures from this scaling are likely understood as a stochastic forcing that results in 'large' amplitude jumps on a 'short' time scale over the extent of the resonant well (29) that effectively destroys the phase velocity - group velocity resonance (Polzin and Lvov 2017), parallel to the cleanly articulated Markov approximation for the one-dimensional model captured in (26). The diagnostics presented in Section 3 provide the where-with-all to assess this.

*c. The Closure Problem*

The advective contribution in the ensemble average transport equation (14) changes the no-flux character obtained with the kinetic equation, as we now have a rational for the GM76 internal wave spectrum to support turbulent mixing by supplying energy for a wave breaking process. However, the predicted supply rate (31) is an order of magnitude too large.

In order to interpret why we have arrived at this end result, we find it useful to engage in a high level discussion of the generic closure problem, motivated by, for example, Orzag (1973) and Holloway and Hendershott (1977), and in text books, e.g. Lesieur (1997). In the context of Hamilton's equation for the time evolution of $a_\mathbf{p}$, one multiplies by $a_\mathbf{p}^*$, multiplies the complex conjugate of Hamilton's equation by $a_\mathbf{p}$, subtracts the two equations and then averages to obtain an evolution equation for the second-order wave action $n_\mathbf{p} = a_\mathbf{p}^* a_\mathbf{p}$ in terms of the third-order correlation function, e.g. Lvov et al. (2012). The process continues iteratively, deriving an equation for the third-order correlation function that involves fourth-order correlations, building up a hierarchy of unclosed equations. If we eliminate all subscripts, coefficients, and summations, the structure can be schematically represented as

$$d\langle\phi\phi\rangle/dt = \langle\phi\phi\rangle + \langle\phi\phi\phi\rangle \tag{32a}$$

$$d\langle\phi\phi\phi\rangle/dt = \langle\phi\phi\phi\rangle + \langle\phi\phi\rangle\langle\phi\phi\rangle + \langle\phi\phi\phi\phi\rangle^C \tag{32b}$$

$$d\langle\phi\phi\phi\phi\rangle/dt = \langle\phi\phi\phi\phi\rangle + \langle\phi\phi\rangle\langle\phi\phi\phi\rangle + \langle\phi\phi\phi\phi\phi\rangle^C \tag{32c}$$

$$\ldots \quad ,$$

in which the superscript $C$ denotes the non-reducible cumulant. The intent of a closure is to truncate the hierarchy.

In (decaying) turbulence, the right hand side of (32b) reads $r.h.s = \langle\phi\phi\rangle\langle\phi\phi\rangle + \langle\phi\phi\phi\phi\rangle^C$. Discarding the fourth-order cumulant in this equation is referred to as the Quasi-Normal (QN) approximation. This is not a statement that the statistics of turbulence are Gaussian, rather it is a statement that the fourth-order cumulant can be neglected for all times in comparison to the remaining terms. This approximation leads to a prediction of negative energy in the energy containing range of the turbulent spectrum (Ogura 1963). Orzag (1973) addressed this by proposing the fourth-order cumulant be approximated as a linear damping term in the third-order equation. This approximation is referred to as the Eddy Damped Quasi-Normal (EDQN) approximation: eddy energy (a second-order moment) is not damped, it is the third-order moments that represent energy exchange that are damped. A final approximation, that the damping time scale varies on a time scale much longer than the time over which $\langle\phi\phi\rangle\langle\phi\phi\rangle$ evolves, represents a Markov approximation, and one obtains the Eddy Damped Quasi-Normal Markovian (EDQNM) approximation.

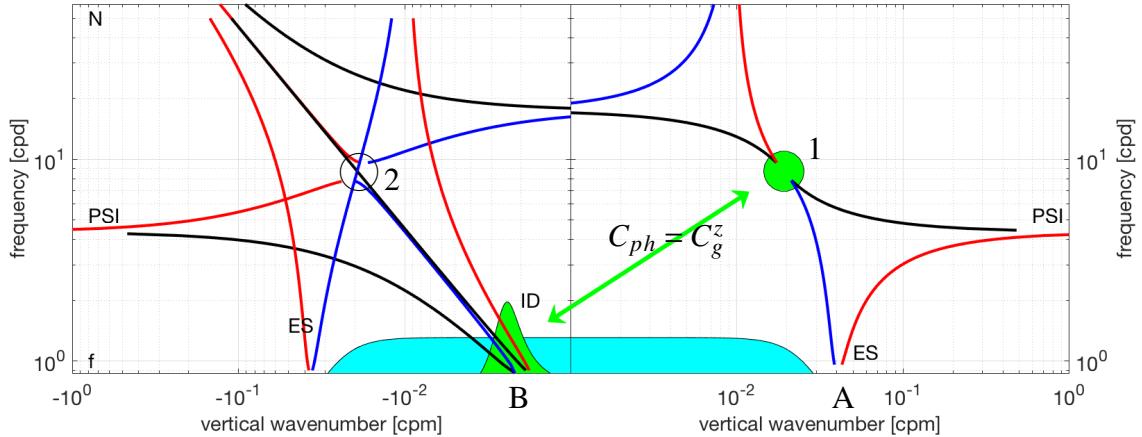

Figure 9. The resonant manifold of the internal wave problem in the situation where the three horizontal wavevectors are either parallel or anti-parallel, plotted in a vertical wavenumber - frequency space, for a wave at the center of the green circle. With rotation, extreme scale separations in horizontal wavenumber lead to Bragg scattering (ES) and a phase velocity $C_{ph}$ equals group velocity $C_g^z$ resonance condition (ID) being located at the Coriolis frequency $f$. This study focuses upon the latter class, with scale separation in both horizontal and vertical wavenumber. Near-resonant ID conditions are depicted in green, bandwidth limited non-resonant ID forcing in cyan. Vertical wavenumber - frequency combinations pertaining to a fourth-order cumulant are labeled $1, 2, A, B$. The Bragg scattering resonance is only scale-separated in horizontal wavenumber.

The parsing of the hierarchy (32) in wave turbulence is slightly different. Equations (32) represent the slow time evolution of wave amplitudes rather than the immediate consequence of linear wave propagation. In the long time limit, the third moments become indefinitely large over a vanishingly small subset of the possible interactions. That subset is the resonant manifold, figure 9. This long time limit leads to a self-consistent description of nearly-resonant interactions, i.e. the broadened kinetic equation (eq. 2.15) and mass operator (eq. 2.17) of LP. The representation obtained by identifying $\langle\phi\phi\phi\rangle$ with $\langle\phi\phi\rangle\langle\phi\phi\rangle$ and substituting in (32a) is the Resonant Interaction Approximation.

It is from this perspective that we can understand Holloway's commentary (Holloway 1980, 1982) in a brighter light. The early work on internal wave kinetic equations (reviewed in Müller et al. (1986)) took the tack of simply deriving the scattering cross-sections and asserting the resonant limit without considering the construction of a broadened kinetic equation. We paraphrase:

1. In constructing a self-consistent kinetic equation one wants a system of field coordinates for which a simple linear combination results in canonical coordinates. Otherwise one needs to express the wave basis in a Taylor series expansion in wave amplitude, and lack of clarity interpreting the broadened equations will ensue. Lagrangian field coordinates (Olbers 1973; McComas 1975; Meiss et al. 1979, not to be confused with ray path coordinates) require a Taylor series expansion about the assumed smallness of the amplitude to arrive at canonical coordinates.

2. Attempts at deriving a self-consistent internal wave kinetic equation from the stand point of a stationary observer are doomed to failure for the same reason that afflicted Kraichnan (1959) in the context of 3-d turbulence: Doppler shifting of the small scales by the large. Kraichnan (1965) deals with this by effectively using only terms in the scattering cross-sections related to pressure and viscosity and thus, in some manner, references Langrangian estimates of correlation time scales.

3. One really wants to know the time scale $\Gamma^{-1}$ (the inverse resonant bandwidth, eq. 2.17 of LP) because closure of the third-order moment equation (32b) ultimately rests upon an assumption that the time integration has converged. Substitution of possible relevant time scales challenge the underpinning assumption that the kinetic equation describes a slow time evolution, but there are only guesses about the relevant time scales. See points (i) and (ii) above.

Item 1 is accomplished in Lvov and Tabak (2004): isopycnal field coordinates lead directly to canonical coordinates. Items 2 and 3 are addressed in Polzin and Lvov (2017). Using a self-consistent kinetic equation for isopycnal coordinates presented in Lvov et al. (2012), Polzin and Lvov (2017) demonstrate that the resonant bandwidth $\Gamma$ suffers from the Doppler shift defect and, moreover, demonstrate that a frequency renormalization does not alter the transport estimate. We therefore conceived of the one-dimensional ray-tracing model (Polzin and Lvov (2017) and Section 2) to investigate. This has led to the articulation of a master equation (14) for the ray path formalism (LP), which in turn over-predicts the finescale parameterization metric for downscale transports by an order of magnitude!

We propose the following two-part interpretation for that over-prediction.

First, action conservation of a single packet and the resulting ensemble average (14) are incomplete expressions of extreme scale-separated dynamics. The ray tracing paradigm reveals a phase locking of high-frequency waves with inertial shear along a phase velocity equals group velocity condition (figure 4) that implies the creation of statistically inhomogeneous conditions in which regions of large inertial shear host accumulations of high-frequency wave stress, figure (4). These conditions are just those that are subject to relaxation by the Elastic Scattering triad, which we present as a Bragg scattering process. In figure 9 we schematically represent the refractive mechanism of 'Induced Diffusion' using waves '1' and 'B'. For given high-frequency wave $\mathbf{p}_1 = (k, 0, m)$, the Bragg scattering mechanism concerns a wave $\mathbf{p}_2 = (k, 0, -m)$ and a low frequency wave at $p = (0, 0, 2m)$, skematicized as transfers between waves '1', '2' and 'A'. The wave stress (momentum flux density) is $u'w' \propto k C_g^z E(\mathbf{p})$. Thus wave 2 has the opposite sign of energy transfer from the inertial wave than wave 1. Bragg scattering therefore damps the accumulation of of wave-momentum by transferring wave energy into another high-frequency wave of opposite sign vertical wavenumber, and hence vertical group velocity, at constant horizontal wavenumber.

In order to promote this from the status of simple conjecture, we point the reader to a step-by-step derivation of ray-tracing and associated action density conservation presented in LP using an extreme scale-separated Hamiltonian structure. The underpinning assumptions are that (i) there are three waves, one of which has much larger amplitude than the others, (ii) that the small amplitude waves have similar frequencies that are both greater than that of the large amplitude wave, and (iii) the small amplitude waves have significantly greater *horizontal* wavenumber than the large amplitude wave. Note that the specification on wave frequency conditions the aspect ratio $k/m$ of the waves. No constraint on vertical wavenumber is implied. Thus, both phase velocity equals group velocity and Bragg scattering resonances in figure 9 are retained. The next step is to derive an evolution equation for the correlation function $N_{1,2} = \langle a(\mathbf{p}_1) a^*(\mathbf{p}_2) \rangle$ in which $a(\mathbf{p})$ are canonical amplitudes at wavenumbers $\mathbf{p}_1$ and $\mathbf{p}_2$. The time evolution of this correlation function can be obtained using Hamilton's equation:

$$i\dot{N}_{1,2} = \int (A_{3,\mathbf{p}_1} N_{3,\mathbf{p}_2} - A_{\mathbf{p}_2,3} N_{\mathbf{p}_1,3}) d3 \,, \tag{33}$$

with wavenumber '3' representing a variable of integration. The factor $A_{\mathbf{p}_1,\mathbf{p}_2}$ represents the Hamiltonian density of the extreme scale-separated resonances as given by rewriting eq. 3.11 of

LP:

$$A(\mathbf{p}_1,\mathbf{p}_2) \cong \sigma_{\mathbf{p}}\delta(\mathbf{p}_1-\mathbf{p}_2) + i\frac{1}{2}\frac{\sigma}{N}\Upsilon_z(\text{A or B}) + \frac{1}{2}\frac{\sigma}{\Pi_0}\hat{\Pi}(\text{A or B}) . \tag{34}$$

This setup applies to both resonances, which we have indicated by replacing the arguments of $\Upsilon_z/N$ and $\hat{\Pi}/\Pi_0$ with (A or B). The factors $\Upsilon_z/N$ and $\hat{\Pi}/\Pi_0$ are Fourier coefficients describing the vertical shear and isopycnal separation variability (what the oceanographic community refers to as 'strain') as vertical gradient analogues of kinetic and potential energy. Ray tracing and action spectral density conservation are obtained by operating on the evolution equation (33) with a Wigner transform and subsequent Taylor series expansion that focusses upon the 'B' half of the spectral domain. Although the Bragg resonance 'A' and the group velocity equals phase velocity resonance 'B' imply evaluation of the Fourier coefficients at vastly different vertical wavenumber magnitudes, the oceanic vertical wavenumber spectrum for these vertical gradients is independent of vertical wavenumber in its power law subrange. Thus the Fourier coefficients at these dissimilar scales have similar magnitudes.

Quantifying the coupling and consequent diminishment of downscale transports implies including the Bragg scattering term while integrating along a ray. Methods for incorporating such effects in an ensemble transport equation exist in literature concerning open quantum systems. The essence of this, an exponential diminuation of the ID coupling, is potentially recoverable using layered-media theory (e.g. Fouque et al. 2007). In the mean time, however, note that one can obtain from the spatially homogeneous kinetic equation the Bragg scattering action balance (McComas and Müller 1981a):

$$\frac{\partial [n(\mathbf{p}_1) - n(\mathbf{p}_2)]}{\partial t} = \tau_r^{-1}[n(\mathbf{p}_2) - n(\mathbf{p}_1)] ,$$

with $\mathbf{p}_1$ and $\mathbf{p}_2$ interpreted as in figure 9. For the GM76 spectrum, the Bragg scattering relaxation time scale $\tau_r$ is equal to the slow 'Induced Diffusion' time scale (McComas and Müller 1981a), which we have identified as $\partial_m D_{33}/m$ and demonstrated to be equal to the drift time scale $\langle \dot{m} \rangle/m$. In short, we intuit that refraction and scattering are exquisitely balanced in this problem. It would represent the ray path formalism's equivalent of an EDQN closure.

Second, the limit of $\tau \to \infty$ as the lower bound of integration in (21) is the signature Markov approximation. Whether such a replacement is reasonable, though, requires justification. The

Table 2. Timescales

Time scale definitions for figure 10.

| | | |
|---|---|---|
| $\Gamma$ | | resonant bandwidth (fast ID; rms Doppler shift) |
| $\tau_c$ | $\frac{\int_{t-\tau}^{t}\langle(\dot{m}(t)-\langle\dot{m}(t)\rangle)(\dot{m}(t-\tau)-\langle\dot{m}(t-\tau)\rangle)\rangle d\tau}{\langle(\dot{m}(t)-\langle\dot{m}(t)\rangle)^2\rangle}$ | correlation timescale |
| $\tau_i$ | $\frac{m}{\langle\dot{m}\rangle}$ | mean drift (slow ID) |
| $\tau_r$ | $\tau_i$ | Bragg scattering |
| $\tau_{\text{event}}$ | $\frac{1}{2}\frac{m}{\dot{m}}\times(3\pi[\frac{\omega}{Nf}]^2 e_0 m_* M_r)^{1/3}$ | $\frac{\text{width of resonant well}}{\text{mean drift projected onto }M}$ |

results of our one-dimensional model at oceanic amplitudes suggest otherwise, but this requires defining the relevant time scale for the sake of comparison. We believe this time scale to be $m/\langle\dot{m}\rangle$ since the wave frequency enters into the lag-autocorrelation time scale (26) and vertical wavenumber and frequency are related through the dispersion relation. This is a by-product of the *ad hoc* construction of a spatial scale separation (3) criterion in our one-dimensional model to supplant a dynamically self consistent specification of the wave packet envelope. On the other hand, if downscale transports are reduced by an order of magnitude by Bragg scattering, which is entirely reasonable considering the disparity between the prediction (31) and community wisdom articulated in Polzin et al. (2014), then an extended one-dimensional closure could fit within the domain of an EDQNM scheme.

A schematic ordering of relevant time scales at oceanic amplitudes is contained in table 2 rendered in figure 10. After averaging over mesoscale eddy time scales, we represent the long time variations in action spectral density as seasonal. Spectral transports occur within a phase velocity - group velocity resonance that has an event timescale $\tau$event which we estimate as the ratio of the resonant bandwidth $\gamma_M$ (29) and the projection of the average drift rate $\langle\dot{m}\rangle$ onto the inertial field, $\dot{M}_r$, with $\dot{M}_r/M_r = 2\langle\dot{m}\rangle/m$, see Section 2. Mean drift $\tau_i$ and Bragg scattering $\tau_r$ operate on identical time scales. At oceanic amplitudes, these are somewhat shorter time scales than those characterizing resonant events. A time scale separation implies the creation of local conditions that are vertically anisotropic and prone to relaxation by Bragg scattering. At oceanic amplitudes, mean drift rates are comparable to the correlation time scale $\tau_c = ss/\langle\omega\rangle$ for non-resonant forcing and challenge a Markov approximation upon which a diffusive closure is predicated. The bandwidth of the self consistent kinetic equation is $\Gamma$.

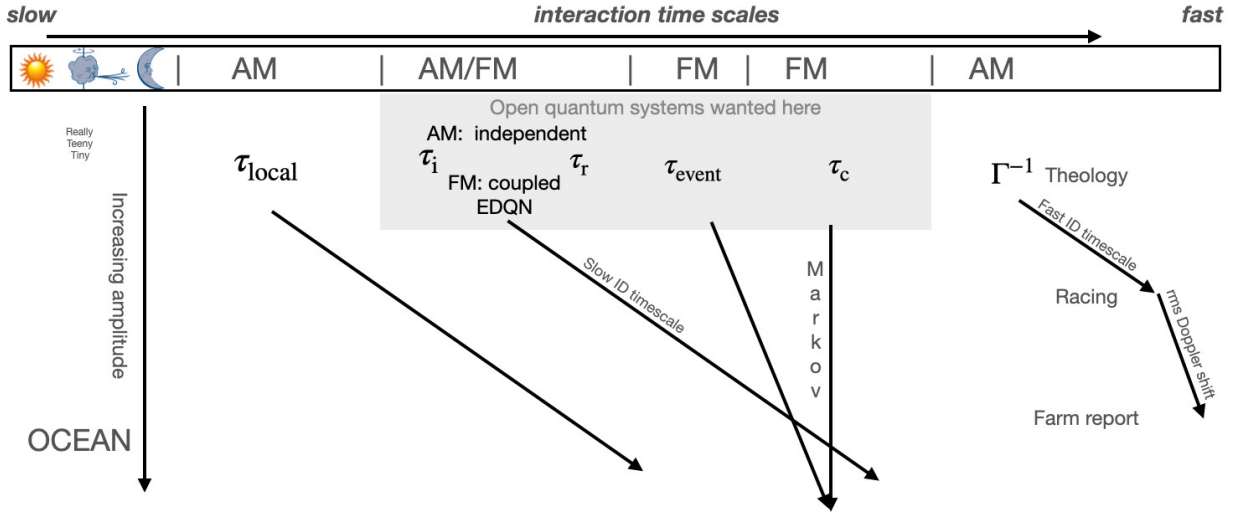

Figure 10. A schematic ordering of time scales. Time scales decrease from left to right and wave amplitude increases from top to bottom. The shortest time scale is associated with the resonant bandwidth $\Gamma^{-1}$ and can be identified as the fast ID time scale of kinetic theory. At oceanic amplitudes this bandwidth degenerates into the rms Doppler shift. The phase velocity equals group velocity $\tau_i$ and Bragg scattering resonances $\tau_r$ operate on similar times scales and can be identified as the slow ID time scale of kinetic theory. These interaction time scales decrease with increasing wave amplitude. A correlation time scale $\tau_c$ associated with non-resonant forcing provides the shortest time scale for the ray tracing simulation. At oceanic amplitudes the interaction time scales are similar to the correlation time scales and provide an issue for the closure of ensemble averaged transports. At oceanic amplitudes the breadth of the phase velocity equals group velocity resonance in the spectral domain is such that ensemble average transit times through the resonance $\tau_{event}$ are longer than the drift time scale $\tau_i$. The longest depicted timescale relates to climatological patters of forcing. The characterization of AM denotes the amplitude-modulated basis of kinetic theory, FM applies to the frequency-modulated paradigm of ray tracing (Polzin and Lvov 2017). The time scale $\tau_{local}$ represents local interactions in the kinetic equation (Dematteis et al. 2022)

We aware of a rigorous proof (Deng and Hani 2021) that fourth-order cumulants are subleading order terms in the expansion (32) under weak linearity for spatially homogeneous systems. Our efforts demonstrate that there are fundamental differences in the statistics of the resonances accumulated along ray paths rather than as a stationary observer. This line of argument is further sustained by the finding in Polzin and Lvov (2017) that the resonant bandwidth $\Gamma$ tends to the rms Doppler shift at finite amplitude while the underpinning dynamics of ray tracing are to represent the variations in the Doppler shift. Individually, Bragg scattering and inertial phase velocity - internal wave group velocity resonances are, in the ray path analysis, the leading order extreme scale-separated processes and have time scales shorter than those associated with 'local' interactions that we intuitively characterize as having a dimension greater than one. Our proposition is that these leading order extreme scale-separated processes are coupled with a resulting diminishment of transports. We draw upon 3-d turbulence to present this coupling as a fourth-order cumulant in the context of an Eddy Damped Quasi-Normal Markov (EDQNM) closure. We regard this as a physically reasonable justification of the 1-d system in the face of the rigorous 3-d proof in Deng and Hani (2021).

Finally, we note that there is an indirect analogy concerning the propagation of electrons in a lattice known as 'Anderson localization'. As the impurities in that lattice become greater, electrical resistance increases in proportion. When the impurities reach a density of two per unit wavelength of the electron, the material suddenly becomes an electrical insulator. Our assessment of the internal wave problem has the same character. A path integral assessment leads us to a description that predicts downscale energy transfer an order of magnitude greater than supported by observations. We forward the hypothesis that inclusion of Bragg scattering physics associated with a background inertial wavefield at half the vertical wavelength of the high-frequency internal wave will similarly shutdown mean drifts to higher wavenumber. Localization occurs in many different physical systems (Lagendijk et al. 2009) with similar concerns about the dimensionality of the system impacting the potential for localization (e.g. Sheng and van Tigglen 2007).

## 6. Summary

As we look back over the landscape of this endeavor, what we have is a well established metric for ocean mixing (Polzin et al. 2014) which, prior to Dematteis et al. (2022), did not have a first principles support. At best, the Finescale Parameterization was underpinned by a heuristic description as an advective spectral closure (Polzin 2004a) in the context of an energy transport equation that eschews action conservation. Application of classical Wave Turbulence to extreme scale-separated interactions arrives at a Fokker-Planck equation expressing wave action diffusion that predicts no downscale spectral transports in vertical wavenumber (6). This result is a consequence of the fact that the GM 3-d action spectrum is independent of vertical wavenumber in its high wavenumber power law regime: there are no gradients of action to support a diffusive transport. Application of ray path techniques in LP arrives at a combined advection / diffusion transport equation (14) that supports downscale vertical wavenumber transports of energy and action for the GM spectrum. In this paper we utilize an idealized representation of high-frequency oceanic internal waves propagating in a background of inertial waves to obtain concrete closures for the ray path formulation. These closures recover the wave turbulence kinetic equation diffusivity and identify the mean drift of wave packets as the gradient of the kinetic equation diffusivity in vertical wavenumber. This represents qualitative progress in reconciling observations with theory. However, predictions for energy transport associated with the mean drift are an order of magnitude larger than the observations. Ray tracing simulations are conducted to assess these closures. There is a tendency for the mean drift to be reduced from these small amplitude scalings at oceanic amplitude, but this reduction is insufficient to ameliorate the glaring order of magnitude discrepancy between prediction and observation. There is a wealth of information available in the ray tracing simulations concerning phase locking and correlation times scales that enable us to effectively order the time scales of a cumulant hierarchy.

Meditation upon this hierarchy in the context of 3-d turbulence produces some parallels. The distinction between wave turbulence and ray-path techniques invites a comparison between Eulerian (Kraichnan 1959) and Lagrangian (Kraichnan 1965) formulations of 3-D turbulence, with Eulerian formulations being prone to contamination by Doppler shifting. The wave kinetic equation represents nonlinear interactions as an amplitude modulation of spatially infinite plane waves and predicts a very rapid adjustment of a spike inserted into an otherwise smooth spectrum (McComas 1977). At oceanic amplitudes, this adjustment time scale tends to an aphysical rms Doppler shift (Polzin and Lvov 2017). The ray-path derivation invokes a Wigner transform that integrates the evolution of spectral spike with its interaction partners, and explicitly represents *variations* in Doppler shifting as the underpinning dynamics. In 3-D turbulence, the change from Eulerian to Lagrangian perspectives changes predicted power laws from $k^{-3/2}$ and $k^{-5/3}$. In the internal wave problem, the concept of ensemble averaging wave packets following ray trajectories provides motivation to include a mean drift term in the transport equation. This is a qualitative difference, and revised estimates of downscale transport overshoot the mark by an order of magnitude. These quantitative differences are potentially resolved by parallels concerning the role of fourth-order cumulants.

In 3-D turbulence, fourth-order cumulants are understood to provide a systematic damping of downscale transports associated with third-order terms (Orzag 1973). Here we propose a Bragg scattering process that reduces the downscale transports associated with a phase-velocity / group velocity resonance. We interpret Bragg scattering as a fourth-order cumulant playing the role of a third-order damping. We find that Bragg scattering and the mean drift have essentially identical time scales.

We identify the Bragg scattering process in terms of the time evolution of the correlation between the two high-frequency wave amplitudes having similar horizontal wavenumber and oppositely signed vertical wavenumber. Incorporating such effects into a wave-packet action balance and deriving a corresponding transport equation are subjects of current research. Similarly, the dynamics that control the envelope structure of the wave packet and interactions of a wave packet with the residual flow associated with the packet's envelope structure (Bühler and McIntyre 2005) have been discarded in the ray tracing paradigm (Gershgorin et al. 2009). Accounting for the latter requires loosening the specification of layer-wise constant potential vorticity (Lvov and Tabak 2004) which is a key piece of isopycnal coordinates representing a canonical Hamiltonian system. This, again, is a topic of current research.

*Acknowledgments.* KP and YL (Grant #N000141712852) gratefully acknowledge support from the Office of Naval Research for this project.
*Data availability statement.* No data were created for this paper.